\documentclass[12pt,preprintnumbers,aps,amssymb,nofootinbib,amsmath]{revtex4}
\usepackage{epsfig}
\usepackage{graphicx}

\def\eq#1{{(\ref{#1})}}
\def\fig#1{{Fig.~\ref{#1}}}

\newcommand{\beq}{\begin{equation}}
\newcommand{\eeq}{\end{equation}}
\newcommand{\beqar}[1]{\begin{eqnarray}\label{#1}}
\newcommand{\eeqar}{\end{eqnarray}}

\newcommand{\im}{\mathrm{Im}}

\newcommand{\arcsinh}{\mathrm{arcsinh}}
\newcommand{\arctanh}{\mathrm{arctanh}}
\begin{document}

\preprint{}

\title {From Color Glass Condensate to Quark Gluon Plasma\\
through the event horizon
}

\author{Dmitri Kharzeev}

\affiliation{ Nuclear Theory Group, \\ Physics Department, Brookhaven National Laboratory,\\
Upton, NY 11973-5000, USA}

\author{Kirill Tuchin}
 
\affiliation{ Nuclear Theory Group, \\ Physics Department, Brookhaven National Laboratory,\\
Upton, NY 11973-5000, USA}

\date{\today}
\preprint{BNL-NT-05/2}

\begin{abstract} 
We propose a new thermalization scenario for heavy ion collisions which 
at sufficiently high energies implies the phase transition to the quark--gluon plasma. 
The key ingredient of our approach is the Hawking--Unruh effect: 
an observer moving with an acceleration $a$ experiences the influence of a thermal bath with an effective 
temperature $T = a / 2\pi$, similar to the one present in the vicinity of a black hole horizon. 
For electric charges moving in external 
electromagnetic fields of realistic strength, the resulting temperature appears too small to be detected. 
However for partons   
in strong color fields the effect should be observable:
in the Color Glass Condensate picture, the strength of the color-electric field is $E \sim Q_s^2/g$ ($Q_s$ is the 
saturation scale, and $g$ is the strong coupling), the typical acceleration is $a \sim Q_s$, and the heat bath temperature 
is $T = Q_s / 2\pi \sim 200$ MeV. In nuclear collisions at sufficiently high energies the effect can induce  a rapid thermalization over the time period of $\tau \simeq 2\pi/Q_s \simeq 1\ {\rm fm}$ accompanied by phase transitions. We consider a specific 
example of chiral symmetry restoration induced by a rapid deceleration of the colliding nuclei. We argue that parton saturation in the initial nuclear wave functions is a necessary pre--condition for the formation of quark--gluon plasma. We discuss the implications of our "black hole thermalization" scenario for various 
observables in relativistic heavy ion collisions.

\end{abstract}
\maketitle

\newpage
\section{Introduction}

 In 1974 Hawking \cite{Hawking:1974sw} demonstrated 
that black holes evaporate by quantum pair production, and behave as if they have 
an effective temperature of 
\beq
T_H = {\kappa \over 2 \pi},
\eeq
where $\kappa = (4 G M)^{-1}$ is the acceleration of gravity at the surface of a black hole of mass $M$; $G$ is the Newton constant.  The thermal 
character of the black hole radiation stems from the presence of the event horizon, which hides 
the interior of the black hole from an outside observer.  
The rate of pair production in the gravitational background of a black hole can be evaluated 
by considering the tunneling through the event 
horizon. Parikh and Wilczek \cite{Parikh:1999mf} showed that the imaginary part of the action for this classically forbidden 
process corresponds to the exponent of the Boltzmann factor describing the thermal emission\footnote{Conservation laws also imply a non-thermal correction to the emission rate  \cite{Parikh:1999mf}, possibly causing a leakage of information from the black hole.}.

 Unruh \cite{unruh} has found that a similar effect arises in a uniformly accelerated 
frame, where an observer detects an apparent thermal radiation with the temperature 
\beq \label{unruhtemp}
T_U = { a \over 2 \pi};
\eeq
($a$ is the acceleration). The event horizon in this case emerges due to the existence of causally disconnected regions of space--time \cite{Lee:1985rp}, conveniently described by using the Rindler coordinates.  

\vspace{0.3cm}

The effects associated with a heat bath of temperature (\ref{unruhtemp}) usually are not easy to detect because 
of the smallness of the acceleration $a$ in realistic experimental conditions. For example, for the acceleration of gravity 
on the surface of Earth $g \simeq 9.8 \ {\rm m\ s^{-2}}$ the corresponding temperature is only $T \simeq 4 \times 10^{-20}  \ 
{\rm K}$. A much larger accelerations can be achieved in electromagnetic fields, and Bell and Leinaas \cite{Bell:1982qr} considered the possible manifestations of the Hawking--Unruh effect  in particle accelerators. They argued that the presence of an apparent heat bath can cause beam depolarization. Indeed, if the energies of spin--up $E_{{\uparrow}}$ and spin--down $E_{{\downarrow}}$ states of a particle in the magnetic field of an accelerator differ by $\Delta E = E_{{\uparrow}} - E_{{\downarrow}}$, the Hawking--Unruh effect would lead to the 
thermal ratio of the occupation probabilities
\beq\label{depol}
{N_{{\uparrow}} \over N_{{\downarrow}}} \simeq \exp\left(-{\Delta E \over T_U}\right),
\eeq
where the Unruh temperature \eq{unruhtemp} is determined 
by the particle acceleration. According to \eq{depol}, a pure polarization state 
of a particle in the accelerator is inevitably diluted by the acceleration. 

If the energy spectrum of an accelerated observer is continuous, as is the case 
for a particle of mass $m$ with a transverse (with respect to the direction 
of acceleration) momentum $p_T$, a straightforward extension of \eq{depol} leads to a 
thermal distribution in the "transverse mass" $m_T = \sqrt{m^2 + p_T^2}$:
\beq\label{tranmass}
W_m(p_T) \sim \exp\left( - {m_T \over T_U} \right).
\eeq
\vskip0.3cm
An important example is provided by the dynamics of charged particles in external electric fields. 
Since a particle of transverse mass $m_T$ and charge $e$ in an external electric field of strength $E$ 
moves with an acceleration $a = e E / m_T$, the thermal distribution \eq{tranmass} 
with the temperature $T_U = a / 2 \pi$ is given by
\beq\label{tranmassel}
W^E_m(p_T) \sim \exp\left( - {2 \ \pi  m^2_T \over e E} \right), 
\eeq    
which differs from the classical Schwinger result \cite{schwinger} for the momentum distribution 
of particles produced from the vacuum only by a factor of two in the exponent. 
We will establish in Section \ref{pairprod} that this interpretation of Schwinger's result 
indeed holds, and will show how a more careful treatment avoids the factor of 2 
discrepancy. 
Note that the result \eq{tranmassel} cannot be expanded in powers 
of the field strength $E$, and thus cannot be reproduced in perturbation theory. 
This fact has a geometrical interpretation in the accelerated frame: the Bogoliubov transformation relating 
the particle creation and annihilation operators in Minkowski and Rindler spaces 
describe a rearrangement of the vacuum structure which cannot be captured by 
perturbative series. 

\vspace{0.3cm}

Perhaps the largest accelerations accessible to experiment at present are achieved 
in the collisions of relativistic heavy ions. Indeed, at RHIC accelerator at BNL, 
the heavy ions collide with c.m.s. momenta of $100 \ {\rm GeV}$ per nucleon, 
and the strong interactions of their parton constituents lead to the production 
of a sizable fraction of final state partons at rest in the c.m.s. frame. This happens 
over a very short time $\Delta t < 1 \ {\rm fm}$, so a typical deceleration is $a \simeq (\Delta t)^{-1} \simeq 200 \ {\rm MeV}$ (we will find later that due to the presence of strong color 
fields  in the initial state described by parton saturation the achieved deceleration is even higher,  $a \sim 1\ {\rm GeV}$). Such deceleration 
should lead to observable effects, which include an apparently fast 
thermalization of the produced partonic state. Indeed, the experimental data from RHIC  \cite{Arsene:2004fa, Adcox:2004mh, Back:2004je, Adams:2005dq}  
and their analysis in terms of the hydrodynamical approach  \cite{Gyulassy:2004zy, Shuryak:2004cy, Teaney:2004qa, Heinz:2004ar,Hirano:2004rs} suggest that thermalization 
occurs over a time of about  $1 \ {\rm fm}$, much faster than is expected on the 
basis of perturbative calculations \cite{Molnar:2001ux}.

In this paper we will argue that  the Hawking--Unruh effect indeed can cause 
rapid thermalization in relativistic heavy ion collisions. If we describe 
the initial parton wave function of the nucleus in the Color Glass Condensate picture \cite{GLR, Mueller:wy, Blaizot:nc, MV, YuK, jmwlk, ILM, Levin:1999mw, KLN}  
where the strength of the color-electric field is $E \sim Q_s^2/g$ ($Q_s$ is the 
saturation scale, and $g$ is the strong coupling), the typical acceleration achieved in a collision is $a \sim Q_s$, and the resulting heat bath temperature 
is $T = Q_s / 2\pi \sim 200$ MeV. As will be discussed below, in nuclear collisions at sufficiently high energies and/or sufficiently large 
atomic number, the effect can induce  a rapid thermalization over the time period of $\tau \simeq 2\pi/Q_s$ accompanied by phase transitions. 

\vspace{0.3cm}

The paper is organized as follows: In section \ref{pairprod} we consider the case 
of particles moving in strong external fields of various kind, and discuss the 
correspondence between the Schwinger and Hawking--Unruh effects. 
We show in particular that a pulse of a strong (color)electric field produces thermally  distributed charged particles. 
In section \ref{secthoriz} we formulate an approach to relativistic heavy ion collisions 
which takes into account the Hawking--Unruh effect. Soft partons in this picture 
are produced by tunneling through the event horizon in Rindler space, and emerge 
thermally distributed with temperature  $T = Q_s / 2\pi \sim 200$ MeV. 
We derive the notion of a critical acceleration $a_c$ within the dual string picture, and show that it is related 
to the Hagedorn temperature $T_{Hagedorn}$ by the relation $T_{Hagedorn} = a_c / 2 \pi$. We argue that parton saturation in the initial wave function is a necessary 
condition of deconfinement in the final state, since it allows to create decelerations 
larger than $a_c$, and thus to produce a thermal medium with a temperature $T > T_{Hagedorn}$. As an explicit example of 
a phase transition caused by a strong deceleration $a \sim Q_s$, we discuss in section \ref{phasetrans} 
the restoration of chiral symmetry in the framework of Nambu-Jona-Lasinio model 
formulated in Rindler space. We then discuss the implications of our scenario for 
various observables in relativistic heavy ion collisions. We conclude the paper with 
the Summary, where we discuss the limitations of our approach, its relation to other 
developments, and suggest some directions for future studies.

\section{Pair production in strong external fields: tunneling through the event horizon  }\label{pairprod}
 
 \subsection{Schwinger pair production as a Hawking--Unruh phenomenon}

In a classic 1951 paper \cite{schwinger}, Schwinger considered the 
dynamics of QED in the strong field domain.
 He constructed an effective action describing the coupling of electromagnetic fields 
 to charged particles; in an electric field the action acquired 
an imaginary part corresponding to the pair creation. For future discussion it is instructive to rederive here the result of Schwinger in the Euclidean time formalism. 
 Let us consider the action of a charged particle of mass $m$ in a constant external electric field $E$:
\beq\label{actfield}
S\,=\,\int\,(\,-\,m\,ds\,-\,e\,\varphi\,dt\,)\,, 
\eeq
where $\varphi$ is the electric potential.  In a constant electric field $E$ the electric potential is $\varphi=-Ex$ modulo an additive constant. Equations of motion of this particle are 
\beq\label{eqmotEM}
\frac{d p_x}{dt}\,=\,e\,E\,,\quad \frac{d p_\bot}{dt}\,=\,0\,.
\eeq
The velocity $\vec v$ of the particle is $\vec v = d\vec x/dt=\vec p/\mathcal{E}_\mathrm{kin}$, where the kinetic energy of the particle is $ \mathcal{E}^2_\mathrm{kin}=m^2+\vec p^2$.  With the initial conditions $\vec p(0)=0$, $\vec x(0)=0$ equations of motion \eq{eqmotEM} can be integrated yielding the trajectory
\begin{subequations}
\beq\label{traja}
v(t)\,=\,\frac{a\,t}{\sqrt{1\,+\,a^2\,t^2}}\,,
\eeq
\beq\label{trajb}
x(t)\,=\,a^{-1}\,(\,\sqrt{1+a^2\,t^2}\,-\,1)\,,
\eeq
\end{subequations}
where we can identify the constant $a$ with acceleration of the particle:
\beq\label{accel}
a\,=\,\frac{d}{dt}\,\frac{v}{\sqrt{1\,-\,v^2}}\,=\,\frac{eE}{m}\,;
\eeq
the trajectory \eq{trajb} is shown by the upper curve in Fig.1.

At $t<0$ the particle is coming from $x\rightarrow 
\infty$ gradually losing its kinetic energy in favor of the potential 
one and finally stopping at $x=0$ at $t=0$. This is the turning point of the 
classical trajectory. At $t>0$ the particle accelerates again to the velocity 
$v\rightarrow 1$ at $x\rightarrow\infty$. The action associated with the 
particle moving along the trajectory \eq{trajb} is given by \eq{actfield}. Using $ds^2=(1-v^2(t))\,dt^2$ and substituting the trajectory \eq{traja} and \eq{trajb} into \eq{actfield} we get 
\begin{eqnarray}\label{act}
S(\tau)&=&\int^\tau\,dt\,(-\,m\,\sqrt{1\,-\,v(t)^2}\,+\,e\,E\,x(t))\,\nonumber\\
&=& -\,\frac{m}{a}\,\arcsinh(a\,\tau)\,+\,
\frac{e\,E}{2\,a^2}\,\left(a\,\tau\,(\sqrt{1\,+\,a^2\,\tau^2}\,- 2)\,+\,\arcsinh(a\,\tau)\right)\,+
\,\mathrm{const}\,.
\end{eqnarray} 
In classical mechanics the equations \eq{traja},\eq{trajb} completely specify 
the motion of a uniformly accelerating particle moving under the influence of a constant force 
$\vec F = -e\nabla \varphi $. In contrast, in quantum 
theory the particle has a finite probability to be found under the 
potential barrier $V(x) =eE x$ in the classically forbidden region. Mathematically, it 
comes about since the action 
\eq{act} being an analytic function of $\tau$ has an imaginary part 
\beq\label{impart}
\im \,S(\tau)\,=\,\frac{m\,\pi}{a}\,-\,\frac{e\,E\,\pi}{2\,a^2}\,=\,\frac{\pi\,m^2}{2\,e\,E}\,.
\eeq 
The imaginary part of the action \eq{act} corresponds to the motion of a particle 
in Euclidean time. Substituting $t\rightarrow -it_E$ in \eq{trajb} we find 
the Euclidean trajectory
\beq\label{euctraj}
x(t_E)\,=\,a^{-1}\,(\,\sqrt{1\,-\,a^2\,t_E^2}\,-\,1\,)\,.
\eeq
Note that unlike in Minkowski space the Euclidean trajectory is bouncing 
between the two identical points $x_a=-a^{-1}$ at $t_{E,a}=-a^{-1}$ and 
$x_b=-a^{-1}$ at $t_{E,b}=a^{-1}$, and the turning point $x_a = 0$ at $t_{E,a}=0$ -- 
see the lower curve in Fig.1.  Using \eq{act} we can find the 
Euclidean action between the points $a$ and $b$; it is given by $S_E(x(t_E))=\pi m^2/2 e E $. It 
is well known that a quasi-classical exponent describing the decay of a metastable 
state is given by the Euclidean action of the bouncing solution (see e.\ g.\ \cite{rubakov}), \eq{gamma1}.   
The rate of tunneling under the potential barrier in the quasi-classical approximation is thus given by 
\beq\label{gamma1}
\Gamma_{V\rightarrow m}\,\sim\,e^{-2\,\im S}\,=\,e^{-\frac{\pi\,  m^2}{e\,E}}\,.
\eeq
Equation \eq{gamma1} gives the probability to produce a particle and its antiparticle (each of mass $m$) out of the vacuum by a constant electric field $E$. The ratio of the probabilities to produce states of masses $M$ and $m$ is then
\beq\label{ratprob}
\frac{\Gamma_{V\rightarrow M}}{\Gamma_{V\rightarrow m}}\,=\,
e^{-\frac{\pi\,  (M^2\,-\,m^2)}{e\,E}}\,.
\eeq 

The relation \eq{ratprob} allows a dual interpretation in terms of both Unruh and Schwinger effects (see e.g. \cite{Parentani:1996gd, 
 Gabriel:1999yz, Narozhny:2003ux} and references therein). First, consider a detector with quantum levels $m$ and $M$ moving with a constant acceleration in the constant electric field. Each level is accelerated differently, however if the splitting is not large, $M-m\ll m$ we can introduce the average acceleration of the detector 
\beq\label{averacc}
\bar a\,=\,\frac{2\,e\,E}{M\,+\,m}\,.
\eeq  
Substituting \eq{averacc} into \eq{ratprob} we arrive at 
\beq\label{ratun}
\frac{\Gamma_{V\rightarrow M}}{\Gamma_{V\rightarrow m}}\,=\,e^{\frac{2\,\pi\,(M\,-\,m)}{\bar a}}\,.
\eeq
This expression is reminiscent of the Boltzmann weight in 
a heat bath with an effective temperature (\ref{unruhtemp}): $T = \bar a/2 \pi$. It implies that the detector is effectively immersed in a photon heat bath at temperature $T\approx e   E / \pi m$. This is the renown Unruh effect \cite{unruh}.

\vspace{0.3cm}

Schwinger mechanism traditionally is viewed as a process of creation of virtual particle--antiparticle pairs from vacuum. In the case of electrodynamics this corresponds to the tunneling  of an electron from the Dirac sea \cite{sauter,cnn}. In this process the electron energy changes 
from $\varepsilon_-$ to $\varepsilon_+$, where 
\beq\label{vareps}
\varepsilon_\pm\,=\,\pm\,\sqrt{p^2(x)\,+\,m^2}\,+\,|e|\,E\,x\,.
\eeq
When the quasiclassical action is evaluated along the true trajectory it can be viewed as a function of the final coordinate (or time) of the moving particle. In this case    
its imaginary part  is given by $\im\,S=\int\,dx\,|p(x)|$ where the integral must be evaluated in the region of imaginary $p$ between turning points $a$, $b$ such that $p(x_{a,b})=0$, i.\ e.\  $x_{a,b}=(\varepsilon\pm m)/|e| E$. A simple calculation yields the result for the tunneling probability which coincides with \eq{gamma1}.


\begin{figure}[htb]\label{figtraj1}
\includegraphics[width=12cm]{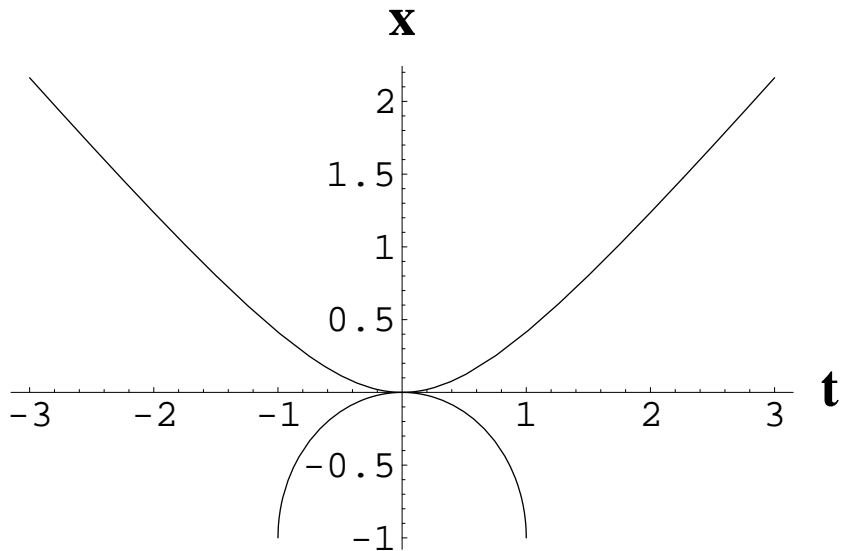}
\caption{Trajectories of an accelerated particle in Minkowski and Euclidean space; the spatial coordinate $x$ and the time 
$t$ are plotted in units of the inverse acceleration, $a^{-1}$.  The upper curve 
represents the Minkowski trajectory of a particle under the influence of a constant force $F$: the particle approaches from $x = + \infty$, 
slows down and stops at $x = 0$, $t = 0$, then turns around and accelerates again towards $x = + \infty$. The trajectory 
is a hyperbola in Minkowski space, or a line with constant value of the coordinate $\rho^2 = x^2 - t^2 = a^{-2}$ in Rindler space.  
The lower curve is the instanton--like Euclidean trajectory of the particle under the potential barrier $V = - F x$: the particle bounces 
between $x = -1/a$ and $x = 0$ with a period of motion $\beta = 2 \pi /a$. The inverse of $\beta$ has a meaning 
of an effective Unruh temperature $T_U = \beta^{-1}$. }
\end{figure}
\vskip0.5cm

Another interpretation of \eq{ratun} can be given if we recall that the uniformly accelerated detector in Minkowski space is equivalent to the inertial detector in the Rindler space. The vacuum in Minkowski space is related to the vacuum in the Rindler space by a non-trivial  Bogoliubov transformation  which shows that the Rindler 
vacuum is populated with the thermal radiation of temperature $T=2\pi/a$ 
in accordance with \eq{ratun}. In general, according to the 
Equivalence Principle of General Relativity, equations of motion in an 
accelerated frame are equivalent to a motion in a certain gravitational 
field. The thermal spectrum  appears then as a result of tunneling through the 
event horizon. The most famous example of such a tunneling process in 
quasi-classical quantum gravity  is Hawking radiation \cite{Hawking:1974sw} where the 
thermal quanta are emitted from the black hole horizon. We would like to note that the methods of General Relativity have been applied 
recently to the studies of different aspects of QCD by several authors \cite{Polchinski:2000uf, Policastro:2001yc, Kharzeev:2002rp, Shuryak:2003ja}.

The method we used for derivation of the Schwinger formula yields another 
important well-known result: constant magnetic field cannot produce pairs 
from vacuum. Indeed, an electron in this case moves with a constant  absolute value of the 
acceleration on a circular orbit. In such a motion 
$v^2\,=\,\mathrm{const}$ and hence $\im S\,=\,0$.

\subsection{Pair production by a pulse of electric field}\label{pulse}

In the situations we consider in this paper the fields can be assumed approximately spatially homogeneous. However, the interaction time is short, so the fields are time-dependent. 
Let us therefore use the method described above to calculate the rate of 
pair production by a spatially homogeneous but time-dependent electric field 
\beq
\vec E=E\,\hat x \cosh^{-2}(t/\mathfrak{t}) .
\eeq
 Our choice for the functional time dependence 
of the field pulse is motivated by the fact that Klein-Gordon and Dirac equations in this case are exactly integrable (see e.g. \cite{Dunne:1998ni}), 
and we can test the accuracy of our quasi-classical result. 

The equations of motion of an electric charge $e$ in this case are given by 
\beq\label{eqmo1}
\dot p_x\,=\,\frac{e\,E}{\cosh^2(t/\mathfrak{t})}\\,\quad \dot p_y\,=\,0\,.
\eeq
With the initial condition $\vec p(0)=0$ they are integrated to give 
\beq\label{eqmo2}
p_x(t)\,=\, e\,E\,\mathfrak{t}\,\tanh(t/\mathfrak{t})\,,\quad p_y(t)\,=\,0\,.
\eeq
The velocity of the particle is given by $\vec v=\vec p/\mathcal{E}_\mathrm{kin}=
\vec p/\sqrt{m^2+p_x^2}$.  Therefore, according to \eq{act}, the kinetic part of the action along the trajectory is 
\beq\label{tact}
S_K(\tau)\,=\, \frac{-\,m^2\,\mathfrak{t}}{\sqrt{m^2\,+\,e^2\,E^2\,\mathfrak{t}^2}}\,
\arctanh\frac{\sqrt{m^2\,+\,e^2\,E^2\,\mathfrak{t}^2}\,\tanh(t/\mathfrak{t})}{\sqrt{m^2\,+\,e^2\,E^2\,\mathfrak{t}^2
\,\tanh^2(t/\mathfrak{t})}}\,. 
\eeq
 The imaginary part of the action \eq{tact} is 
 \beq\label{imp}
\im\, S_K(\tau)\,=\,\frac{\pi\,m^2\,\mathfrak{t}}{\sqrt{m^2\,+\,e^2\,E^2\,\mathfrak{t}^2}}\,.
\eeq

The contribution of the dynamical part to the action is given by the second term in the right-hand-side of \eq{actfield}.  To find an explicit form of the trajectory $x(t)$ let us introduce a new variable $\xi=\sinh(t/\mathfrak{t})$. Then, 
\begin{eqnarray}\label{traj2}
x(\xi)&=&e\,E\,\mathfrak{t}^2\,\int\frac{d\xi\,\xi}{\sqrt{1\,+\,\xi^2}\,
\sqrt{m^2\,+\,\xi^2\,(m^2\,+\,e^2\,E^2\,\mathfrak{t}^2)}}\,\nonumber\\
&=&\frac{e\,E\,\mathfrak{t}^2}{\sqrt{m^2\,+\,e^2\,E^2\,\mathfrak{t}^2}}\,
\ln\left\{\sqrt{1\,+\,\xi^2}\,+\,\sqrt{\,\xi^2\,+\,
\frac{m^2}{m^2\,+\,e^2\,E^2\,\mathfrak{t}^2}}\,\right\}\,+\,\mathrm{const}\,.
\end{eqnarray}
Correspondingly, the action is 
\begin{eqnarray}\label{accc}
&&S_I(\tau)\,=\,e\,E\,\int^\tau x(t)\,dt\,= \nonumber\\
&& \frac{(e\,E\,\mathfrak{t})^2}{\sqrt{m^2\,+\,e^2\,E^2\,\mathfrak{t}^2}}\,
\int^\tau dt\,
\ln\left\{\cosh(t/\mathfrak{t})\,+\,\sqrt{\,\sinh^2(t/\mathfrak{t})\,+\,
\frac{m^2}{m^2\,+\,e^2\,E^2\,\mathfrak{t}^2}}\,\right\}\,+\,S_0(\tau),
\end{eqnarray}
where $S_0(\tau)$ is a certain real entire function of $\tau$. There are two important limits of \eq{accc}. At $\mathfrak{t}\rightarrow\infty$, we can expand the integrand of \eq{accc} to the terms $\sim 1/\mathfrak{t}$. Then the  imaginary part of the action $S_I$ reduces to  $\im S_I(\tau)=\pi m/2e E$ which coincides with the corresponding result in the constant field.

 In the opposite limit $\mathfrak{t}\rightarrow 0$ 
\beq\label{short}
S_I(\tau)\,=\,\frac{(e\,E)^2\,\mathfrak{t}^3}{m}\int^{\tau/\mathfrak{t}} dt'\,\ln(\cosh t')\,+\,S_0(\tau)\,,\quad \mathfrak{t}\,\rightarrow\,0\,.
\eeq
The imaginary part of \eq{short} is calculated by making transformation to the Euclidean time $t_E$ and integrating in the interval $-\pi/2\le t_E \le\pi/2$, i.\ e.\ between the turning points of the potential.
The result is 
\beq\label{imlong}
\im\,S_I(\tau)\,=\,\frac{(e\,E)^2\,\mathfrak{t}^3}{m} \,\pi\ln 2\,,\quad \mathfrak{t}\,\rightarrow\,0.
\eeq

In the limit $\mathfrak{t}\gg a^{-1}=m/eE$, i.\ e.\ when the field $E$ is quasi-constant 
we have 
 \beq\label{prev}
 \im\,S(\tau)\,=\,\im\,S_K(\tau)\,+\,\im\,S_I(\tau)\,=\,\frac{m^2}{2\,|e|\,E}\,,\quad \mathfrak{t}\,\rightarrow\,\infty\,.
 \eeq
Therefore we recover the previous result \eq{impart}.

In the opposite limit of a short pulse the field $E$  varies rapidly, so that  $\mathfrak{t}\ll a^{-1}$, i.\ e.\  the time $\mathfrak{t}$  is much shorter than the inverse acceleration the particle would have obtained in a constant field. In that case $\im S_K\gg \im S_I$ as is seen from \eq{imp} and \eq{imshort}. Thus,
\beq\label{imshort}
\im S(\tau)\,=\,\pi\,m\,\mathfrak{t}\,,\quad a\,\mathfrak{t}\,\ll\,1.
\eeq
Therefore, the decay rate 
\beq\label{gamshort}
\Gamma_{V\rightarrow m}\,\sim\,e^{-\,2\,\pi\,m\,\mathfrak{t}}\,,\quad a\,\mathfrak{t}\,\ll\,1\,
\eeq
becomes independent of the value of the field $E$ in agreement with the quantum mechanical uncertainty relation. 
The result \eq{gamshort} agrees with the $a\,\mathfrak{t}\,\ll\,1$ limit of the exact result obtained in \cite{Dunne:1998ni}. 
Our quasi-classical derivation suggests that  \eq{gamshort} is a general result for all 
electric field pulses of a typical duration $\mathfrak{t}$. We have verified by explicit calculations for several functional dependencies of $E(t)$ 
that this is indeed the case.

We can see that the distribution \eq{gamshort} is again thermal, with an effective 
temperature of $(2 \pi \mathfrak{t})^{-1}$. This result reflects the fact that the thermal character of the distribution is a general property of an accelerated reference frame, 
and is not an artifact of a particular assumption about the field which causes 
the acceleration.  
 
\subsection{Pair production in a chromo-electric field}

The results presented above can be generalized to the case of a color particle moving in a homogeneous chromoelectric field. To this end let us consider the Wong equations \cite{Wong} governing the classical motion of such a particle
\begin{subequations}
\beq\label{wonga}
m\,\ddot x^\mu\,=\,g\,F^{a\,\mu\nu}\,\dot x_\nu\,I_a\,,
\eeq
\beq\label{wongb}
\dot I^a\,-\,g\,f_{abc}\,\dot x^\mu\, A_\mu^b\,I^c\,=\,0\,,
\eeq
\end{subequations}
where a dot denotes the derivative with respect to the proper time and  $\dot x^2=1$. 
The chromoelectric field is given by
\beq\label{chrm}
E^i_a\,\equiv\,F^{i0}_a\,=\,\partial_0\,A^i_a\,-\,\partial_i\,A_a^0\,+\,g\,f_{abc}\,A_b^0\,A_c^i\,.
\eeq
The homogeneous chromoelectric field can be described by the potential 
\beq\label{AA}
A^0_a\,=\,-\,E\, z\, \delta^{a3}\,,\quad A^i_a\,=\,0\,.
\eeq
Substituting \eq{AA} into \eq{wongb} we find that the color isospin vector $I_a$ precesses about the 3-axis with $I_3=\mathrm{const}$. Therefore, \eq{wonga} tells us that 
\beq\label{eqms}
\ddot x\,=\,\ddot y\,=\,0\,,\quad m\,\ddot z\,=\,g\,E\,\dot x^0\,I_3\,.
\eeq
Mathematically, equations \eq{eqms} are equivalent to the equations of motion of an electric charge $g I_3$ in the electric field $E \hat z$. Therefore, all arguments given above are valid for the motion in a homogeneous electric field as well. 
Heavy quark production by Schwinger mechanism in strong color fields within the Color Glass Condensate picture has been previously discussed in 
\cite{Kharzeev:2003sk}.

The constant in time field can also be described with another choice of potentials, which is not related by the gauge transformation to \eq{AA} \cite{Wu:1975vq}. However, that choice leads to the periodic motion of the particle along the $z$-axes \cite{Brown:1979bv} and is of no interest for the physical process we will discuss in the next section.

\section{Event horizon and thermalization in high energy hadronic interactions}\label{secthoriz}

\subsection{Unruh effect, Hagedorn temperature, and parton saturation}\label{hagedorn}

We are now ready to address the case of hadron interactions at high energies, 
which is the main subject of this paper. Consider a high--energy hadron of mass $m$ 
and momentum $P$ which interacts with an external field (e.g., another hadron) and transforms into a hadronic final state of invariant mass $M \gg m$. This transformation is accompanied 
by a change in the longitudinal momentum  
\beq \label{ql}
q_L = \sqrt{E^2 - m^2} - \sqrt{E^2 - M^2} \simeq {M^2 - m^2 \over 2P},
\eeq  
and therefore by a deceleration; we assumed that the particle $m$ is relativistic, with energy $E \simeq p$.

The probability for a transition to a state with an invariant mass $M$ is given by 
\beq \label{probtot}
P(M \leftarrow m) = 2 \pi |{\cal{T}}(M \leftarrow m)|^2 \ \rho(M),
\eeq
where ${\cal T}(M \leftarrow m)$ is the transition amplitude, and $\rho(M)$ is the density of hadronic final states\footnote{$\rho(M)$ should not be confused 
with the Rindler coordinate $\rho$ used in other sections of the paper.}. 
According to the results of the previous section, we expect that under the influence 
of deceleration $a$ which accompanies the change  of momentum (\ref{ql}), 
the probability $|{\cal T}|^2$ will be determined by the Unruh effect and will be given by
\beq \label{occfac}
 |{\cal T}(M \leftarrow m)|^2 \sim \exp(-2 \pi M/a)
\eeq
in the absence of any dynamical correlations; we assume $M \gg m$.

To evaluate the density of states $\rho(M)$, let us first 
use the dual resonance model (see e.g. \cite{dealfaro}, \cite{Satz:1973zc}), in which 
\beq \label{degfac}
\rho(M) \sim \exp\left({4 \pi \over \sqrt{6}}\  \sqrt{b} \ M\right),
\eeq  
where $b$ is the universal slope of the Regge trajectories, related to the 
string tension $\sigma$ by the relation $\sigma = 1/(2\pi b)$.  

The unitarity dictates that the sum of the probabilities (\ref{probtot}) over all finite states $M$ should be finite. 
Therefore, by converting the sum into integral over $M$ one can see that the eqs (\ref{occfac}) 
and (\ref{degfac}) impose the following bound on the value of acceleration $a$:
\beq \label{hag}
{a \over 2 \pi} \equiv T \leq {\sqrt{6} \over 4 \pi} \ {1\over \sqrt{b}} \equiv T_{Hag}.
\eeq  
The quantity on the r.h.s. of (\ref{hag}) is known as the Hagedorn temperature \cite{Hagedorn:1965st} -- the 
"limiting temperature of hadronic matter" derived traditionally from hadron thermodynamics. 
In our case it stems from the existence of a "limiting acceleration" $a_0$:
\beq \label{limac}
a_0 = \sqrt{3 \over 2} \ b^{-1/2}.
\eeq 
 
The meaning of the "limiting temperature" in hadron thermodynamics is well-known: above it, 
hadronic matter undergoes a phase transition into the deconfined phase, in which the quarks and gluons 
become the dynamical degrees of freedom. To establish the meaning of the limiting acceleration (\ref{limac}), let us 
consider a dissociation of the incident hadron into a large number $n \gg 1$ of partons. In this case the phase space density (\ref{degfac}) 
can be evaluated by the saddle point method ("statistical approximation"), with the result (see e.g. \cite{keijo})
\beq \label{rhostat}
\rho(M) \sim \exp( \beta M); 
\eeq
where $\beta^{-1}$ is determined by a typical parton momentum in the center-of-mass frame of the partonic 
configuration. When interpreted in partonic language, eq(\ref{degfac}) thus implies a constant value of mean transverse 
momentum ${\bar p}_T \sim \beta^{-1} \sim b^{-1/2}$. On the other hand, in the parton saturation picture, 
the mean transverse momentum has to be associated with the "saturation scale" $Q_s$ determined by the 
density of partons in the transverse plane within the wave function of the incident hadron (or a nucleus). 
This leads to the phase space density $\log\rho(M) \sim M/Q_s$. The unitarity condition and the formulae 
(\ref{occfac}), (\ref{probtot}) thus lead us to the acceleration $a \sim Q_s$, which can exceed (\ref{limac}), and 
to the conclusion that the final partonic states are described by a thermal distribution with the temperature 
$T\sim Q_s/(2\pi)$.
The same result can be obtained by considering the acceleration (\ref{accel}) $ a = g E / m$ of a parton with off-shellness $m \equiv \sqrt{p^2} \simeq Q_s$ in an external color field $gE \simeq Q_s^2$. Using Wong equations \eq{eqms} we derive $a=Q_s$ and  
\beq \label{tempsat}
T = {Q_s \over 2 \pi}.
\eeq  

It is interesting to note that to exceed the limiting acceleration 
(\ref{limac}), and thus the limiting Hagedorn temperature (\ref{hag}) for the produced hadronic matter, one has to build up 
strong color fields, exceeding $gE_0 \sim 1/b$. This is achieved by parton saturation in the Color Glass Condensate, 
when $gE \simeq Q_s^2 > gE_0$ at sufficiently high energies and/or large mass numbers of the colliding nuclei. 
Parton saturation in the initial wave functions thus seems to be a necessary pre-requisite for the emergence of 
thermal deconfined partonic matter in the final state. We will discuss this conjecture in more detail below.   
  
According to arguments in the previous sections, the thermal distribution is built over the time period of 
\beq
t_\mathrm{therm} \simeq T^{-1} = {2 \pi \over Q_s}. 
\eeq  
As discussed above, this apparent thermalization originates from the presence of the event horizon in an accelerating frame:  
the incident hadron decelerates in an external color field, which causes the emergence of the causal horizon. 
Quantum tunneling through this event horizon then produces a thermal final state of partons, in complete analogy 
with the thermal character of quantum radiation from black holes.  

\subsection{The space--time picture of relativistic heavy ion collisions}

\begin{figure}[htb]
\noindent
\vspace{-0.3cm}
\begin{minipage}[b]{.46\linewidth}
\includegraphics[width=8.3cm]{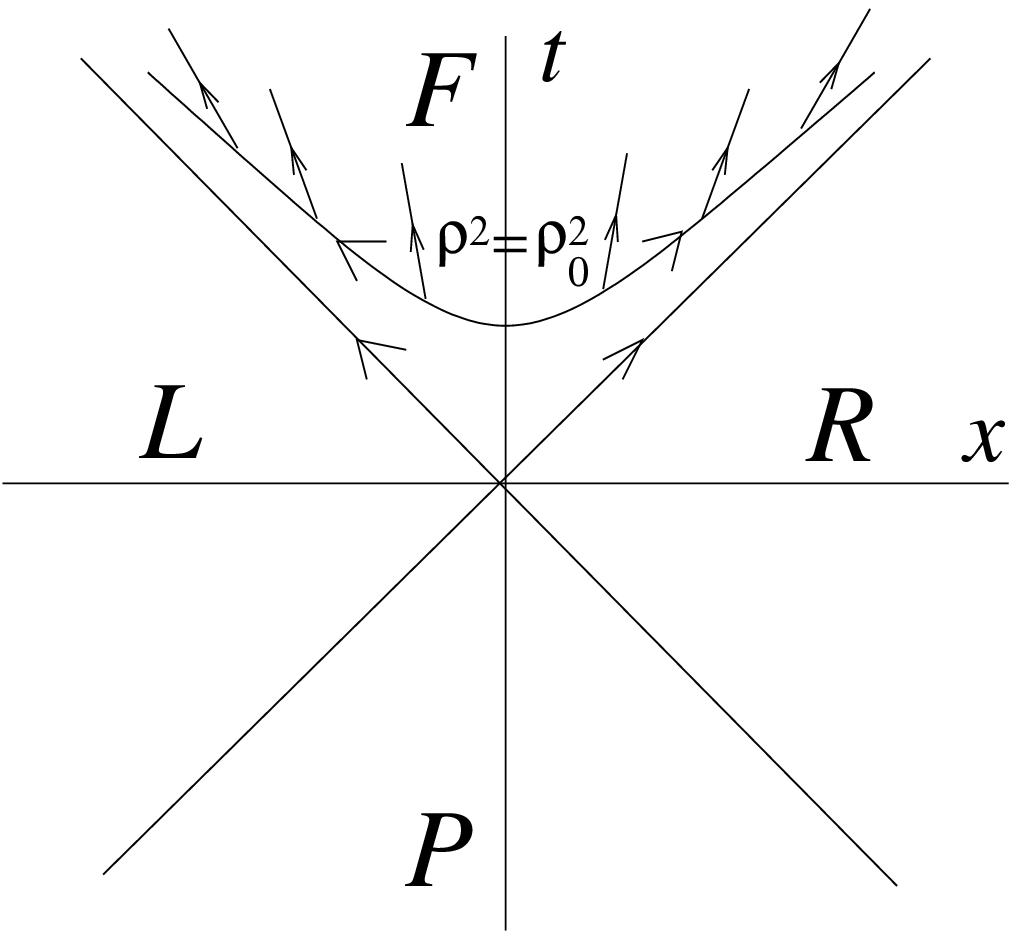}
\end{minipage}\hfill
\begin{minipage}[b]{.46\linewidth}
\includegraphics[width=8.3cm]{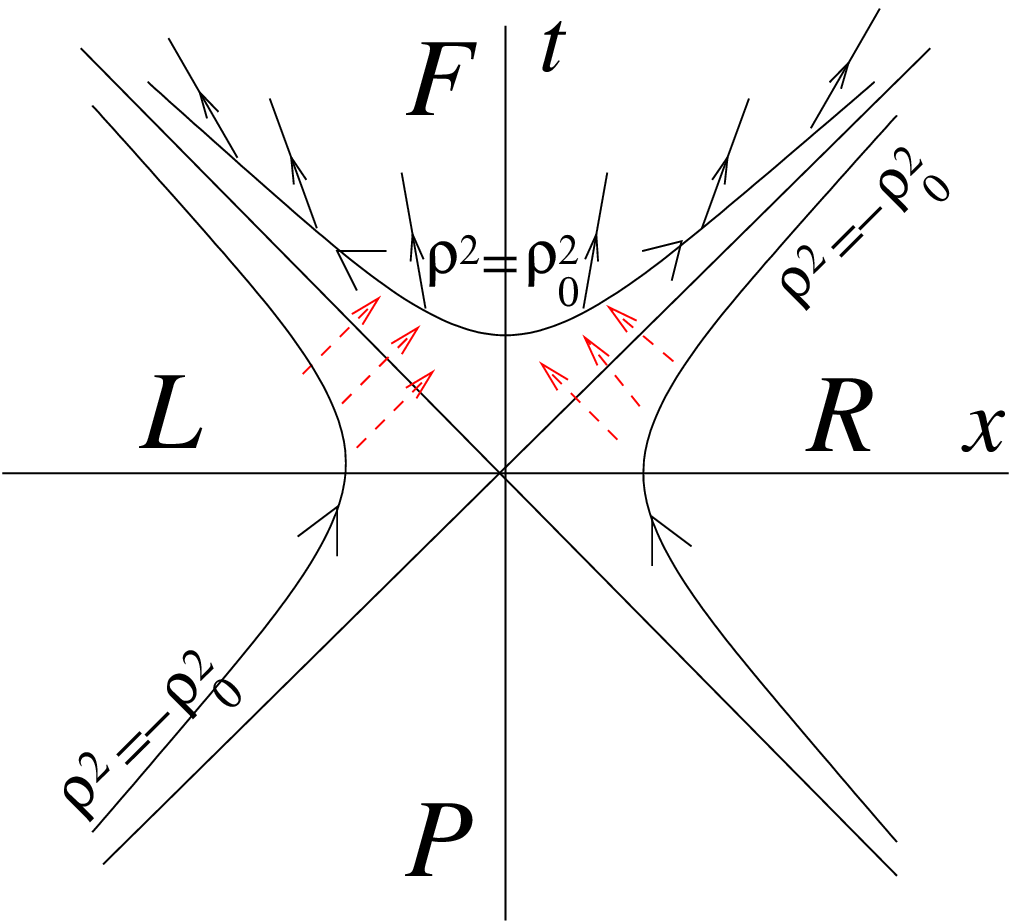}
\end{minipage}
\caption{
The space--time picture of a relativistic heavy ion collision. Left: heavy ions approach the interaction region 
around $x = t = 0$ along the light cone from $x = t = - \infty $ and $x = -t = \infty$. The collision at $x = t = 0$ 
produces the final state particles which approach the mass shell at some proper time $\tau = (t^2 - x^2)^{1/2}$ 
(or, equivalently, along the surface of Rindler space $\rho^2 = \rho^2_0 = x^2 - t^2 < 0$).  The produced particles are 
distributed in rapidity, or in the Rindler coordinate $\eta = {1 \over 2} \ln \left|{t + x \over t - x}\right|$.
Right: also shown in the left (L) and right (R) sectors are the trajectories of space-like $p^2 \sim - Q_s^2 < 0$ 
partons confined in the initial nuclear wave functions characterized by the saturation scale $Q_s$. 
Approaching the interaction region around $x = t = 0$ the partons from the colliding nuclei begin to interact, 
which leads to their deceleration with $|a| = Q_s$; the trajectories in the left and right sectors are the hyperbolae with 
$\rho^2 = - \rho^2_0  > 0$, $|a| = Q_s = \rho^{-1}$. For partons in the right (R) sector, the surface $\rho^2 = 0$, $\eta = + \infty$ 
is the event horizon of the future; the information about the left (L) and the future (F) sectors is hidden from them. 
Quantum tunneling of partons from left (L)  and right (R) sectors through this event horizon into the future (F) indicated 
by dashed arrows 
creates a thermal state of parton matter with the temperature $T = Q_s / 2 \pi$.     
}
\label{rindler}
\end{figure}

The conventional space--time picture of a relativistic heavy ion collision is depicted in the left panel of Fig.\ref{rindler}. 
The colliding heavy ions approach the interaction region along the light cone from $x = t = - \infty $ and $x = -t = \infty$. 
The partons inside the nuclei in the spirit of the collinear factorization approach are assumed to have a vanishing 
transverse momentum $k_T$, have a zero virtuality $k^2 = - k_T^2 =0$, and thus are also localized on the light cone 
at $ \pm x = t$. The collision at $x = t = 0$ produces the final state particles with transverse momenta $p_T$ which according 
to the uncertainty principle approach their mass shell at a proper time $\tau = (t^2 - x^2)^{1/2} = \tau_0 \sim 1/p_T$. 

\vspace{1cm}

For further discussion it is convenient to introduce the Rindler coordinates
\beq\label{rcoord}
\rho^2 = x^2 - t^2; \hspace{1cm} \eta = {1 \over 2} \ln \left|{t + x \over t - x } \right|;
\eeq
The surface of a fixed proper time which is a hyperbola in Minkowski space thus represents a line at $\rho^2 = x^2 - t^2 = \rho^2_0 < 0$ in Rindler space. The Rindler coordinate $\eta$ in high energy physics is often called a space--time rapidity. 

Consider now the case when partons in the wave functions of the colliding nuclei have non-vanishing transverse momenta, as in the Color Glass 
Condensate picture where their transverse momenta are on the order of the saturation scale $Q_s$. In this case 
the partons are space--like $k^2 = - k_T^2$ and are located off the light cone. 
As the colliding nuclei approach each other, these partons begin to interact; note that since they are space--like, 
their interactions are acausal, and are responsible for the breakdown of factorization for the parton modes with $k_T \leq Q_s$.  The interactions of partons with the color fields $g E \simeq Q_s^2$ of another nucleus decelerate them, with a typical acceleration $|a| \simeq Q_s$. The space--time trajectories of the partons, according to 
eq.(\ref{trajb}) are thus given by hyperbolae in Minkowski space, or by the lines with a fixed value of $\rho^2 = - \rho^2_0 = Q_s^2 > 0$ in Rindler space. These trajectories are shown on the right panel of Fig.\ref{rindler}. For partons moving in the left (L) 
and right (R) sectors of space--time with an 
acceleration $|a| = \rho^{-1}$ the light cone surfaces $\rho^2 = 0$, $\eta = \pm \infty$ represent the event horizon 
of the future (F). The information from the future is hidden from them, and the sector F is classically disconnected 
from L and R. However, as discussed above in Section \ref{pairprod}, the future sector F can be reached from 
the left L and right R sectors by quantum tunneling. When evaluated in the quasi--classical approximation, 
for a parton of mass $m$ and transverse momentum $p_T$ the tunneling probability is 
\beq\label{thermpart}
W_m(p_T) \sim \exp\left( - {2 \pi \sqrt{m^2 + p_T^2} \over Q_s}\right).
\eeq
Eq.(\ref{thermpart}) describes a thermal distribution in the "transverse mass" $m_T \equiv  \sqrt{m^2 + p_T^2}$ with the 
temperature $T = Q_s / 2 \pi$.  

The partons produced in the sector of future F reach the mass shell at a typical proper time $\tau_0 = (- \rho^2_0)^{1/2} = 1 / Q_s$. Note that the Rindler trajectory $\rho^2 = - \rho^2_0$ of the produced partons in sector F is related to the 
trajectories of colliding partons $\rho^2 = \rho^2_0$ in the sectors L and R by crossing symmetry. The thermalization 
of the produced partons is reached over the time $\tau_{therm} = T^{-1} = 2 \pi / Q_s$.

\section{Implications of the ``black hole thermalization" for relativistic heavy ion collisions}

\subsection{Rapid thermalization}

There is an ample amount of evidence \cite{Arsene:2004fa, Adcox:2004mh, Back:2004je, Adams:2005dq} that heavy ion collisions at RHIC energies lead to a thermalized state of matter (see \cite{Gyulassy:2004zy} and references therein). Hydrodynamical models  \cite{Shuryak:2004cy, Teaney:2004qa, Heinz:2004ar} appear successful in describing the collective flow of the produced 
hadrons, but only if the thermalization time $\tau_{therm}$ (at which the hydrodynamical evolution begins) is short, $\tau_{therm} \leq 1$ fm. None of the existing approaches had so far succeeded in explaining such a
fast thermalization, despite some promising work ranging from the "bottom-up" scenario based 
on parton saturation \cite{Baier:2000sb, Jeon:2004dh} to the consideration of collective plasma instabilities 
\cite{Mrowczynski:1994xv, Randrup:2003cw, Romatschke:2003ms, Arnold:2003rq, Mrowczynski:2004kv, Arnold:2004ti,Rebhan:2004ur}.

The mechanism of "black hole thermalization" proposed in this paper produces soft thermal partons 
with the initial temperature of $T = Q_s / 2 \pi$; the thermalization time in this case is $\tau_{therm} = T^{-1} = 2 \pi / Q_s$. The analysis of experimental data at RHIC in the framework of KLN saturation model \cite{KLN} 
yields for Au-Au collisions the values of $Q_s \simeq 0.9 \div 1.5\ {\rm GeV}$ depending on centrality. 
This corresponds to the temperatures of $T \simeq 140 \div 240\ {\rm MeV}$, and to thermalization times 
of about $\tau_{therm} \simeq 1$ fm. Note that the lower bound on the initial temperature according 
to our discussion in section \ref{hagedorn} is given by the Hagedorn temperature. Indeed,  
even if the saturation in the initial wave function is not reached, the deceleration in hadronic collisions is 
determined by the string dynamics; the acceleration in this case is determined by the string tension, related through the density 
of hadron states to the Hagedorn 
temperature, see \eq{hag}.

\subsection{Phase transitions}\label{phasetrans}

The arguments presented in this paper are admittedly somewhat schematic. Nevertheless, we hope that our approach 
may provide a useful theoretical tool for understanding the dynamics of hadronic interactions. As an explicit example of using the geometry 
of Rindler space we will now address an important issue of phase transitions induced by relativistic heavy ion 
collisions.

As seen from \eq{accel} the stronger the external field, the larger acceleration it causes since $a\propto E$. The strength of the saturated color field of a nucleus increases with the energy of the collision squared $s$ and the atomic number $A$  as $E\propto Q_s^2\sim A^{1/3} s^{\lambda/2}$ where $\lambda\approx 0.3$. Thus the acceleration of particles in hadron (nucleus) -- hadron (nucleus) collisions  grows at higher energies and for higher atomic masses as a power. On the other hand, Unruh effect relates the acceleration to the temperature of the thermal bath of particles surrounding the accelerated one, see \eq{unruhtemp}.
Therefore, by increasing the energy of the collision one increases the temperature of the produced thermal bath, \eq{tempsat}. This allows tuning an order parameter characterizing the thermal bath of the produced particles.  In particular, by varying the saturation scale in a system with spontaneously broken symmetry one can observe the phase transition to the symmetric phase.
Here we would like to discuss an example of such a phase transition; to be concrete, we will concentrate on the chiral symmetry restoration.  

\vspace{0.3cm}

Rindler space is often considered in general relativity as an approximation to the Schwarzschild metric of 
a large black hole; consequently, a number of important calculations had already been performed 
in this space. In this section, we will rely on these results; our treatment will closely follow a recent 
paper by  Ohsaku   \cite{Ohsaku:2004rv}. 

Consider the $N$-flavor Nambu-Jona-Lasinio (NJL) model:
\beq\label{lagNJL}
\mathcal{L}(x)\,=\,\bar\psi(x)\,i\,\gamma^\nu(x)\,\nabla_\nu\,\psi(x)\,+\,\frac{\lambda}{2\,N}[(\bar \psi(x)\,\psi(x))^2\,+\,(\bar \psi\,i\,\gamma_5\,\psi(x))^2]\,,
\eeq
where $\psi$ is the Dirac field and $\lambda$ is the coupling. The gamma matrices satisfy the following anti-commutation relations:
\beq\label{comm}
\{\gamma_\mu(x),\gamma_\nu(x)\}\,=\,2\,g_{\mu\nu}(x)\,.
\eeq
A uniformly accelerated observer will measure the thermal spectrum of $\psi$'s with temperature $T=2\pi/a$.  To find an effective Lagrangian of an accelerated observer we 
transform the NJL  theory to the Rindler coordinates using \eq{rcoord}. The interval in Rindler space takes form 
\beq\label{rindint}
ds^2\,=\,\rho^2\,d\eta^2\,-\,d\rho^2\,-dx_\bot^2\,.
\eeq
With the help of \eq{comm} one can express gamma matrices in the Rindler space through those in the Minkowski space. 

\begin{figure}
\includegraphics[width=18cm]{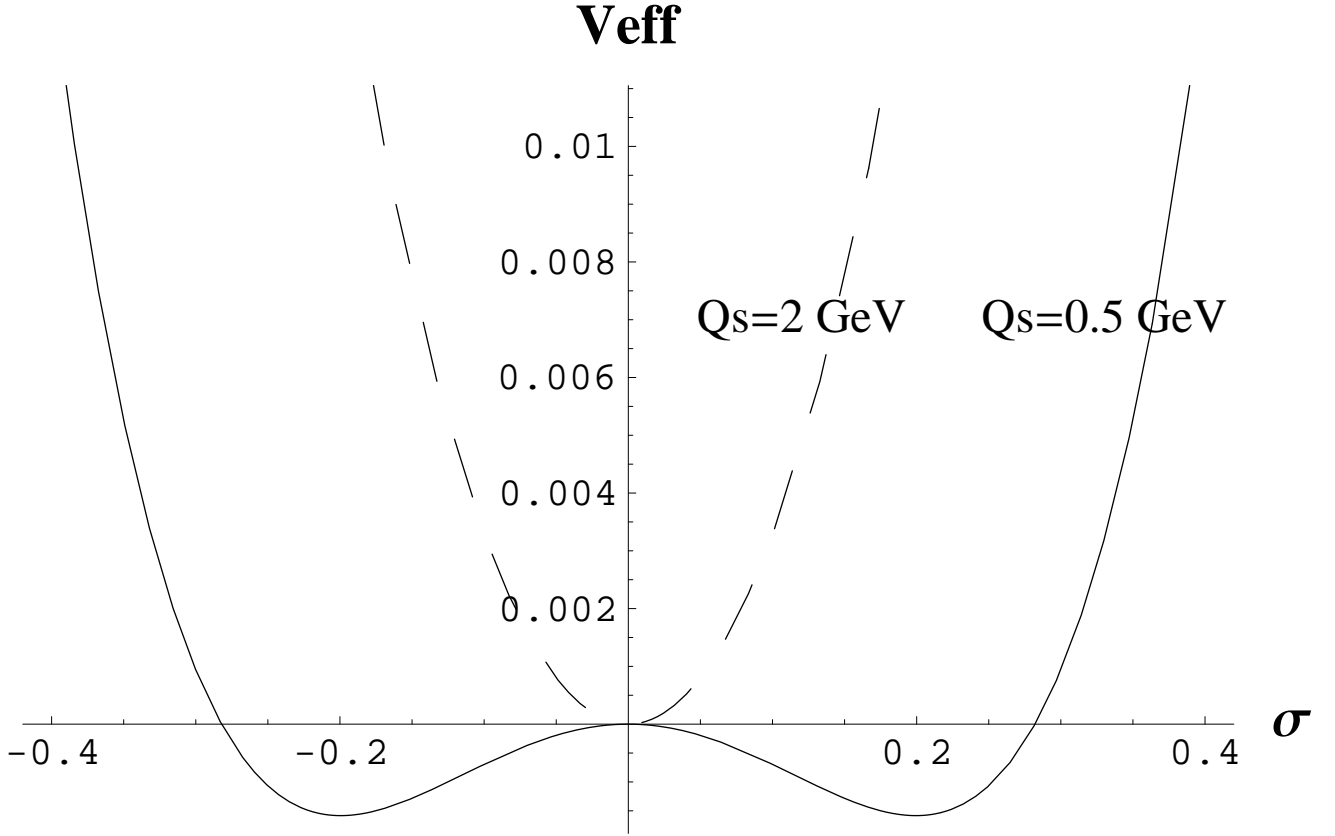}
\caption{The effective potential \eq{veff} of NJL model (scaled by $32\pi$) in an accelerated frame near the critical point \eq{ac}. We chose the critical value of the saturation scale $Q_{s,c}=1$~GeV which is equivalent to the critical temperature of $T_c \simeq 160$~MeV. }\label{fig:veff}
\end{figure}

Introducing as usual the scalar and pseudo-scalar fields as 
\beq\label{sigmapi}
\sigma(x)\,=\,-\frac{\lambda}{N}\,\bar\psi(x)\,\psi(x)\,,\quad \pi(x)\,=\,-\,\frac{\lambda}{N}\,\bar\psi(x)\,i\,\gamma_5\,\psi(x)\,,
\eeq
and performing large-$N$ expansion one can calculate partition function of the problem from which the effective action can be read off:
\beq\label{seff}
S_\mathrm{eff}\,=\,\int d^4x\,\sqrt{-g}\,\left( -\,\frac{\sigma^2\,+\,\pi^2}{2\,\lambda}\right)\,
-\,i\,\ln\det(i\,\gamma^\nu\,\nabla_\nu\,-\sigma\,-i\,\gamma_5\,\pi)\,.
\eeq
This action can be used to derive the gap equation \cite{Ohsaku:2004rv}
\beq\label{gap}
\sigma\,=\,-\,\frac{2\,i\,\lambda\,\sigma}{a}\,\int \frac{d^2k_\bot}{(2\,\pi)^2}\,
\int_0^\infty d\omega\,\frac{\sinh(\pi\,\omega/a)}{\pi^2}\,
\bigg\{\,(K_{i\frac{\omega}{a}+\frac{1}{2}}(\alpha/a))^2\,-\,
(K_{i\frac{\omega}{a}-\frac{1}{2}}(\alpha/a))^2\,\bigg\}\,,
\eeq
where $\alpha^2=k_\bot^2+\sigma^2$. The critical value of the acceleration $a_c$ at which the phase transition occurs is determined by the gap equation \eq{gap} at $\sigma=0$. One gets
\beq\label{ac}
T_c\,=\,\frac{a_c}{2\,\pi}\,=\,\sqrt{\frac{3}{\pi^2}\,\Lambda^2\,-\,\frac{6}{\lambda}}\,,
\eeq
where $\Lambda$ is the cutoff. To study the system near the phase transition $a\approx a_c$ one expands \eq{gap} in the vicinity of the critical point \eq{ac}. The effective potential reads:
\beq\label{veff}
V_\mathrm{eff}(Q _s,\sigma)\,=\,-\frac{1}{96\,\pi^2}\,\big(Q_{s,c}^2-Q_s^2\big)\,\sigma^2\,+\,
\frac{1}{32\,\pi}\,\sigma^4\,+\,\ldots\,,
\eeq
where we used \eq{tempsat}. Eq.~\eq{veff} demonstrates that chiral symmetry restoration occurs whenever the center of mass energy $\sqrt{s}$ or nucleus atomic number $A$ is high enough to ensure that the saturation scale $Q_s$ is larger than the critical value
 $Q_{s,c}=a_c$, see \fig{fig:veff}. 
 
We believe that a similar treatment can be extended to other phase transitions in heavy ion collisions, 
including the deconfinement and the $U_A(1)$ symmetry restoration.

\subsection{Hadron abundances, spectra, and the HBT radii}

It has been known for a long time since the pioneering work by Fermi and Landau \cite{Fermi:1950jd, Landau:1953gs} that statistical models successfully describe many features 
of hadronic reactions at high energies. Indeed, the abundances of produced hadrons by and large follow 
the statistical ones with an effective temperature $T \sim 150$ MeV 
(see \cite{Braun-Munzinger:2003zd, Becattini:1997ii, Becattini:2000jw}  and references therein); moreover, the hadron spectra at 
small "transverse mass" $m_T \equiv \sqrt{m^2 + p_T^2}$ look approximately thermal, $\sim \exp(-m_T/T)$.  
  
On the other hand, it has been found that the variation of the transverse momentum spectra at RHIC with 
centrality (the "transverse flow") exhibits an approximate scaling \cite{Schaffner-Bielich:2001qj, DiasdeDeus:2003fg} in $m_T/Q_s$. 
The picture proposed in this paper provides a natural explanation to this fact: a thermal spectrum \eq{thermpart} with $T = Q_s/2 \pi$ obviously scales in $m_T/Q_s$, even though the shape of the thermal distribution of course can be affected by the hydrodynamical evolution.

Moreover, it has been found both in \cite{KLN} and in \cite{Schaffner-Bielich:2001qj} that the saturation scale 
has to be "frozen" at low energies and/or peripheral $AA$ collisions at a lower bound of about $Q_{s 0} \simeq 0.9$ GeV to describe the data. In view of 
our present discussion, this cutoff is related to the critical acceleration, or the Hagedorn temperature: 
$ Q_{s 0} = a_c = 2 \pi T_{Hagedorn}$. This may explain why the elementary $e^+e^-$ and $p({\bar p})p$ collisions 
exhibit many statistical features \cite{Becattini:1995if, Becattini:1997rv}: the "subcritical" color fields with the strength 
described in the dual resonance model by the string tension  
induce the acceleration $a \simeq a_c = 2 \pi T_{Hagedorn}$.

\vspace{0.3cm}

A well--known puzzle stemming from RHIC results is the short duration of hadron emission as implied 
by the ratio of HBT radii $R_{side}/R_{out} \simeq 1$. This is difficult to achieve if the thermalization is reached 
by the consecutive interactions of the produced particles. We speculate that our scenario with a short production time 
$\tau = 2 \pi /Q_s$ of the apparently thermalized particles may help to solve this problem, but our  
claim of course has to be verified by an explicit calculation. 
  
\section{Summary and discussion}

In this paper we have proposed a new "black hole thermalization" scenario for hadron and heavy ion collisions.
The key idea of our approach is the existence of an event horizon caused by a rapid deceleration of the colliding nuclei. 
The apparent thermalization in this case is a general phenomenon known as the Hawking--Unruh effect: 
an observer moving with an acceleration $a$ experiences the influence of a thermal bath with an effective 
temperature $T = a / 2\pi$, similar to the one present in the vicinity of a black hole horizon. 
In hadron collisions, the deceleration $a_c$ is given by Eq. \eq{hag}, and appears related to the Hagedorn temperature: $a_c = 2 \pi T_{Hagedorn}$. 
To exceed the Hagedorn temperature, and to induce the phase transition(s) to the deconfined and chirally symmetric phase, 
one needs a larger deceleration which can be achieved if the color fields within the colliding hadrons or nuclei grow stronger. 
A mechanism leading to strong color fields in the initial wave functions is at present well--known: it is parton saturation in the 
Color Glass Condensate.  In this picture, the strength of the color-electric field is $E \sim Q_s^2/g$ ($Q_s$ is the 
saturation scale, and $g$ is the strong coupling), the typical acceleration is $a \sim Q_s$, and the resulting heat bath temperature 
is $T = Q_s / 2\pi \sim 200$ MeV. 
Such a large deceleration should induce  a rapid thermalization in nuclear collisions over the time period of $\tau \simeq 2\pi/Q_s \simeq 1\ {\rm fm}$ accompanied by phase transitions. 

We have considered an explicit  
example of chiral symmetry restoration induced by a rapid deceleration of the colliding nuclei. 
On these grounds, we have argued that parton saturation in the initial nuclear wave functions is a necessary pre--condition for the formation of quark--gluon plasma. We have discussed the possible implications of our "black hole thermalization" scenario for various 
observables in relativistic heavy ion collisions, including hadron spectra and abundances, and the HBT radii. 

\vspace{0.3cm}

Our discussion of the Schwinger mechanism  in section \ref{pairprod}   was aimed at the understanding of particle production 
in strong color fields in the framework of the geometrical Hawking--Unruh approach. We have argued that the pair production in this case can be 
understood as a quantum tunneling through the event horizon, similar to the quantum evaporation of black holes. In our case, the event horizon 
appears due to the acceleration of particles in external fields. Schwinger mechanism has been extensively discussed as a model for 
multi-particle production in hadronic and nuclear interactions \cite{cnn, Kerman:1985tj, Cooper:1992hw, Bialas:1999zg}. It was also argued that this pair production can drive the system towards equilibrium 
through successive interactions. Our proposal here is quite different: we show that the spectrum of the produced particles is inherently 
thermal, and that this apparent thermalization is achieved over a very short time of $t_{form} \simeq 2 \pi / Q_s$.  

Numerical simulations of the classical Yang-Mills dynamics in nuclear collisions \cite{Krasnitz:2001qu,Lappi:2003bi} were 
found to lead to approximately thermal transverse momentum spectra of the produced gluons, 
with an effective temperature $T \simeq Q_s/2$, substantially higher than ours.  
In our case, the thermal spectrum arises due to   
a purely quantum effect; since the quantum behavior cannot be reproduced by a classical simulation, 
our results have to be of different physical origin. 

\vspace{0.3cm}

A shortcoming of the quasi-classical approach used in this paper is the necessity to consider the trajectories of particles propagating in external classical 
color fields.  On the other hand, the classical color fields themselves are formed by colored particles -- gluons, with the occupation numbers $\sim 1/ g^2$.
The necessity to treat differently the "particles" and the "classical fields" of course is common for many kinetic approaches. In our case, it may be 
possible to overcome this difficulty by formulating the field theory in Rindler space. We have provided an explicit example of such an approach in section \ref{phasetrans}. 
 

\acknowledgments

This research was supported by the U.S. Department of
Energy under Grant No. DE-AC02-98CH10886.


\end{document}